\documentclass[twocolumn,aps,showpacs,prl]{revtex4}
\usepackage{graphicx,latexsym,endnotes}
\usepackage{psfrag,amsmath}
\usepackage[english]{babel}

\newcommand{\E}{{\mathrm e}}
\newcommand{\arr} {$\rightarrow$}

\newcommand{\media}[1]{\left\langle  #1 \right\rangle}

\newcommand{\pq}[1]{\left[{#1}\right]}

\begin{document}
\author{A. Imparato}
\email{imparato@phys.au.dk}
\affiliation{Department of Physics and Astronomy, University of Aarhus,
  Ny Munkegade, Building 1520, DK--8000 Aarhus C, Denmark}
\author{A. Pelizzola}
\email{alessandro.pelizzola@polito.it}
\affiliation{Dipartimento di Fisica and CNISM, Politecnico di Torino,
  c. Duca degli Abruzzi 24, Torino, Italy}
\affiliation{INFN, Sezione di Torino, Torino, Italy}
\author{M. Zamparo}
\email{marco.zamparo@pd.infn.it} 
\altaffiliation[Present address: ]{Dipartimento di
  Fisica G. Galilei and CNISM, Universit\`a di Padova, v. Marzolo 8,
  Padova, Italy}
\affiliation{Dipartimento di Fisica and CNISM, Politecnico di Torino,
  c. Duca degli Abruzzi 24, Torino, Italy}

\title{Equilibrium properties and force-driven unfolding pathways of RNA molecules}

\begin{abstract}
The mechanical unfolding of a simple RNA hairpin and of a 236--bases
portion of the Tetrahymena thermophila ribozyme is studied by means of
an Ising--like model. Phase diagrams and free energy landscapes are
computed exactly and suggest a simple two--state behaviour for the
hairpin and the presence of intermediate states for the
ribozyme. Nonequilibrium simulations give the possible unfolding
pathways for the ribozyme, and the dominant pathway corresponds to the
experimentally observed one. 
\end{abstract}


\maketitle

The study of the RNA three-dimensional structure has received a boost
by the recent discovery that some RNA molecules act as enzymes in
several key cellular processes in complete absence of protein
cofactors \cite{DounaChec}.

As for proteins, the shape of RNA molecules is strictly connected to
their function, and thus the study of the response to external forces
helps to understand how biomolecules transform mechanical inputs into
chemical signals. Recent experiments and simulations have shown how it
is possible to extract information on the RNA structure by using force
spectroscopy, where RNA molecules are manipulated by using controlled
forces.

In particular, remarkable experimental works \cite{bus1,bus2} have
investigated the connections between the molecular structure of RNA
hairpins and Tetrahymena thermophila ribozyme, and the respective
unfolding pathways under mechanical stress.

Motivated by such experiments, several groups have proposed
theoretical and numerical approaches to the mechanical unfolding of
RNA molecules
\cite{VarieRNA,thiruPnas05,thiruStru06,thiruBJ07,thiruPnas08}.
In particular, by using a coarse--grained G\={o}--model and Molecular
Dynamics (MD) simulations, Thirumalai and coworkers have computed
phase diagrams and free energy landscapes of RNA hairpins
\cite{thiruPnas05,thiruPnas08}, and the unfolding pathways of larger,
more complex RNA molecules \cite{thiruStru06,thiruBJ07}.

However, MD simulations are computationally demanding, and even in the
simple case of the determination of the phase diagram, simulations
have to be restarted for every choice of the model parameters. Here we
introduce a simple discrete model for RNA molecules under external
force, whose thermodynamics is exactly solvable, and as such is able
to provide exact thermodynamical results for any size of the
molecules, in a computation time which is incomparably smaller than
the time needed for simulations. We exploit this model to obtain the
force--temperature $(f,T)$ phase diagram of a small and a large RNA
molecule and their free energy landscape as a function of the
molecular elongation.

Furthermore, by using Monte Carlo simulations (MC), we investigate the
unfolding pathways of the larger molecule, finding that the most
probable path from the native to the unfolded state agrees with the
experimentally determined one. It is worth noting that the present
model has been used to evaluate the phase diagram, the free energy
landscape \cite{AlbPrlJCP07}, and the unfolding pathways
\cite{AlbPrl08} of widely studied proteins, showing a good degree of
agreement with the corresponding experimental results.

{\it The model --}
We use a G\={o} model defined by the energy
\begin{equation}
H(m,\sigma)=-\sum^{N-1}_{i=1}\sum^N_{i < j} \epsilon_{ij} \Delta_{ij} \prod_{k=i}^j m_k - f L,
\label{hamilt}
\end{equation} 
where $L=\sum_{i \le j} l_{ij} \sigma_{ij} (1-m_{i-1}) (1-m_j) \prod_{k=i}^{j-1} m_k$ is the end--to--end length of the molecule and
$m_k = 0, 1$ is associated to the covalent bond between bases $k$ and
$k+1$. $m_k = 1$ (respectively 0) means that this bond is (resp.\ is
not) in a native--like state. Given the state of the $m$ variables,
for an RNA molecule with $N+1$ bases, $n_\sigma(m) = 1 + \sum_{k=1}^N
(1-m_k)$ orientational degrees of freedom are introduced. Such degrees of freedom, $\sigma_{ij}$,
describe the orientation of the $n_\sigma(m)$ native--like stretches,
relative to the external force $f$. Indeed, if a native--like stretch extends from
base $i$ to base $j \ge i$, then $(1-m_{i-1}) (1-m_j)
\prod_{k=i}^{j-1} m_k = 1$. Such a stretch can be as short as a single
base $i=j$. We set as boundary conditions $m_0 = m_{N+1} = 0$. The
orientation $\sigma_{ij}$ of a native--like stretch is also a binary
variable: $\sigma_{ij} = +1$ (respectively $-1$) represents a stretch
oriented parallel (resp.\ antiparallel) to the external force
$f$. $l_{ij}$ is the length of the $i-j$ native stretch, taken from
the Protein Data Base (PDB) and defined as the distance between the phosphorus atoms of bases $i$
and $j+1$. $\Delta_{ij}$ is the element of the contact matrix, which
takes the value 1 if the bases $i$ and $j+1$ are in contact, and 0
otherwise. Within the present model, bases $i$ and $j+1$ are
considered to be in contact if at least two atoms, one from each
base, are closer than $\delta=4$ \AA. Finally $\epsilon_{ij}$ is the
corresponding interaction strength, which is proportional to the
number of atom pairs, which are in contact according to the above
criterion. We have shown \cite{AlbPrlJCP07} that, as far as the
equilibrium thermodynamics is concerned, the sum over the $\sigma$
variables can be performed exactly, and in the $f = 0$ case a
well--known Ising--like model of protein folding is obtained
\cite{VarieWSME}. Stacking interactions are implicitly taken into
account in the present model: one can easily check that, for instance,
the formation of a contact between bases $i+1$ and $j-1$ (if present
in the native structure) is a necessary condition for the formation of
an $i-j$ native contact.
\begin{figure}[h]
\center
\includegraphics[width=8cm]{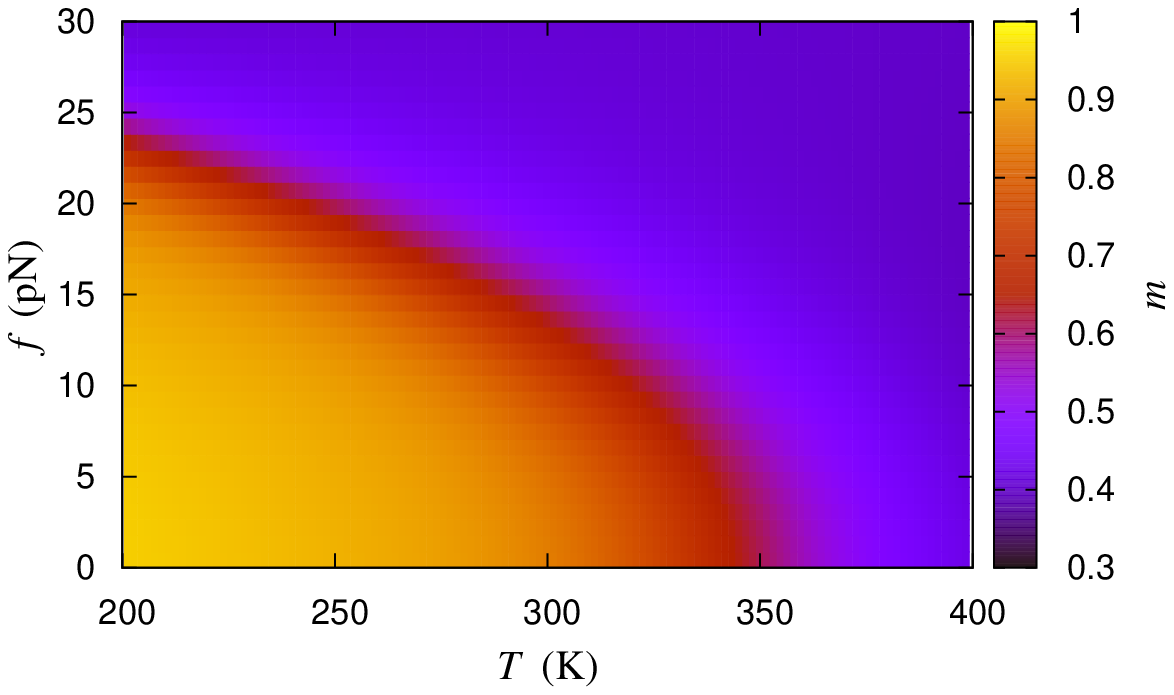}
\includegraphics[width=8cm]{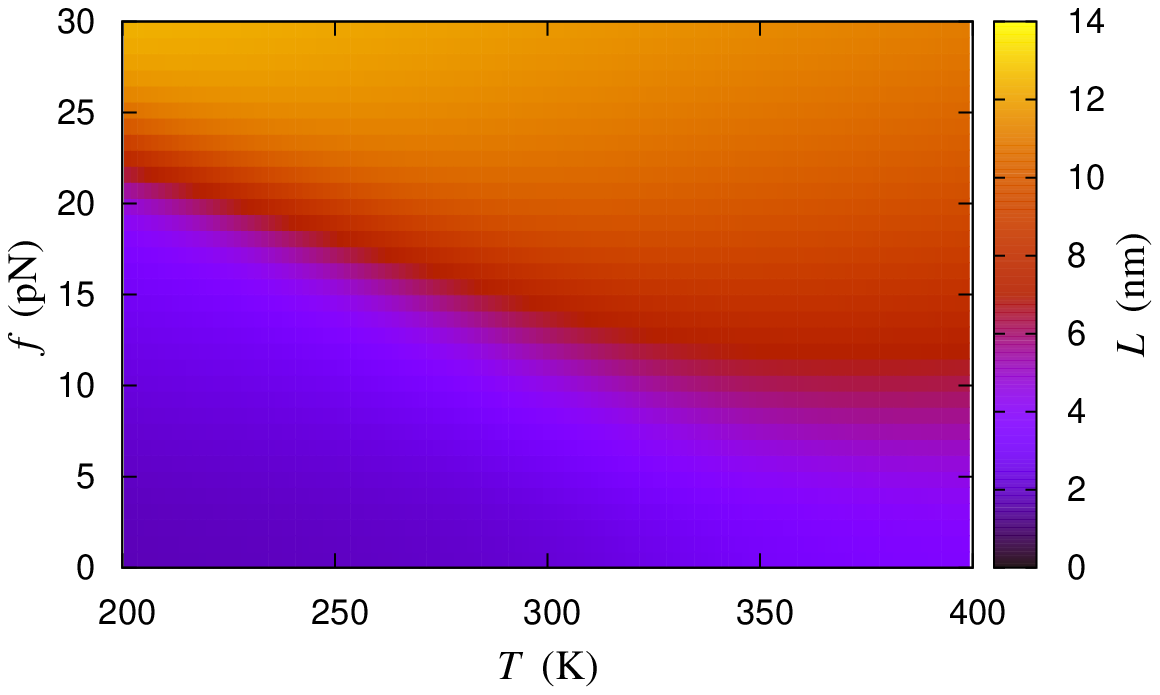}
\caption{(Color online) Force-Temperature phase diagrams of the 1EOR
  molecule. Upper panel: average order parameter $\media m$. Lower
  panel: average end--to--end length $\media L$.}
\label{phase_eor}
\end{figure}

{\it Simple hairpin (P5GA) --} We first consider a 22--nucleotides RNA
hairpin, (PDB code 1EOR, see ref.\cite{append} for the secondary
structure), which is similar to the P5ab in the P5abc domain of group
I intron \cite{bus1}.  We first focus on the equilibrium properties of
the molecule, and then study the unfolding kinetics induced by
external forces.

In fig.~\ref{phase_eor}, the $(f,T)$ phase diagrams are plotted. They show
that P5GA behaves like a two-state system, the transition region from
the folded to the unfolded state being quite narrow. It
is also in good agreement with that by Hyeon and Thirumalai
\cite{thiruPnas05} for the same molecule.

\begin{figure}
\center
\includegraphics[width=8cm]{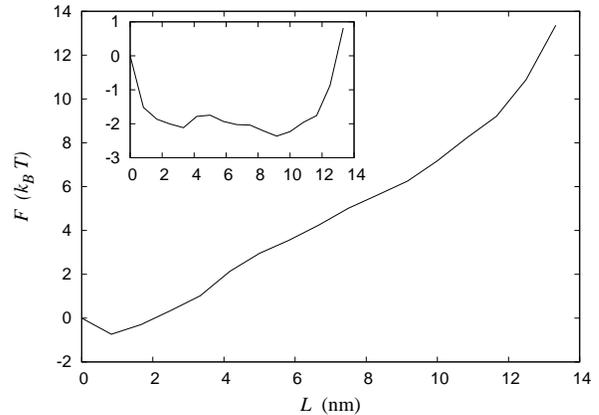}
\caption{Free energy landscape $F_0$ of the 1EOR molecule as a
  function of the molecular elongation $L$. Inset, tilted FEL $F_0-f\,
  L$, with $f=15.4$ pN. }
\label{free_eor}
\end{figure}

An interesting quantity that characterizes the stability of a biopolymer, is the free energy landscape (FEL) $F_0$ as a function of its end-to-end elongation $L$, defined as 
\begin{equation}
F_0(L)=-k_B T \ln\pq{\sum_x \E^{-\beta H_0(x)} \delta (L(x)-L)},
\end{equation} 
where $x$ is the microscopic state of the system, and the sum is
restricted to those states, whose value of the macroscopic
variable $L(x)$ is equal to the argument $L$ of $F_0$.  Here $H_0(x)$ corresponds to the hamiltonian (\ref{hamilt}) with $f=0$. As discussed
in \cite{AlbPrlJCP07}, $F_0(L)$ can be exactly computed in the
present model. The landscape is plotted in
fig.~\ref{free_eor}, for $T=300$ K. When an external constant force
$f$ is applied to the molecule free ends, one gets the tilted
landscape $F(L,f)=F_0(L) -f \, L$, which is plotted in
fig.~\ref{free_eor}, for $f=15.4$ pN. From the phase diagram
in fig.~\ref{phase_eor} one sees that for this value of the
force, the molecule length is about half of its maximum value
$L_{\mathrm{max}}\simeq13.8$ nm. At the same time, the FEL exhibits
two wells at small and large elongation, indicating that the molecule
hops from the folded to the unfolded state, for this value of the
force \cite{thiruPnas05}.
The FEL of the same molecule was obtained in
\cite{thiruPnas05,thiruBJ07,thiruPnas08} by using an off-lattice
coarse grained model and MD simulations. In these papers, $F(L,f)$
was estimated by analyzing the kinetics of the molecule under force,
i.e., by sampling the occupation frequency of those states with a
given elongation $L$. This method is expected to give reliable results
only for small molecules, like the one at issue, where the phase space
of the molecule is sampled according to its equilibrium phase space
distribution.  In the case of larger molecules, at small forces, one
expects that the states with large $L$ are not sampled with the
correct frequency, as computer simulations usually fail to visit rare
states.  On the contrary, the phase diagrams and landscapes, as given by the present model, are exact
results, and can be obtained for any molecule size and any value of
$f$ and $T$.

{\it Tetrahymena thermophila ribozyme (1GRZ) --} In the following we
investigate the thermodynamical equilibrium properties and the
mechanical unfolding of the Tetrahymena thermophila ribozyme
\cite{golden98}, PDB code 1GRZ, whose mechanical unfolding has been
studied both experimentally \cite{bus2} and with computational
techniques \cite{thiruStru06}. We consider the structured part from
base 96 to 331, which exhibits several secondary structure elements
(SSE), named Pn with n=$3,\dots,9$, see ref.~\cite{append}.

\begin{figure}
\center
%\psfrag{ }[ct][ct][1.]{ }
\includegraphics[width=8cm]{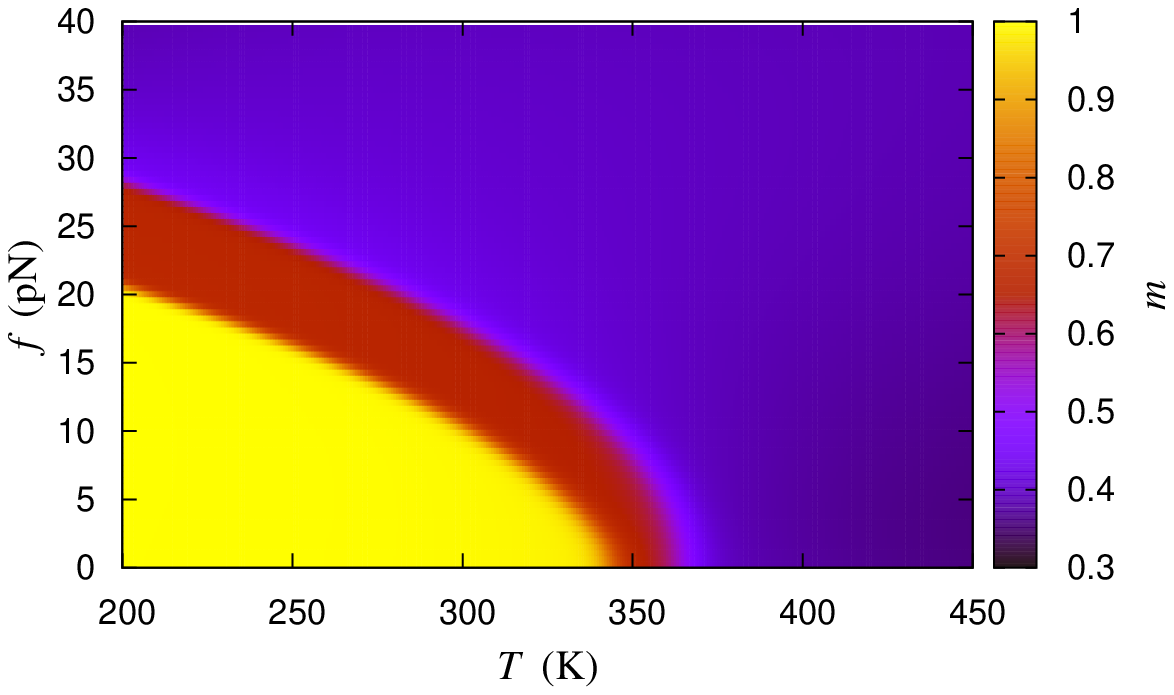}
\includegraphics[width=8cm]{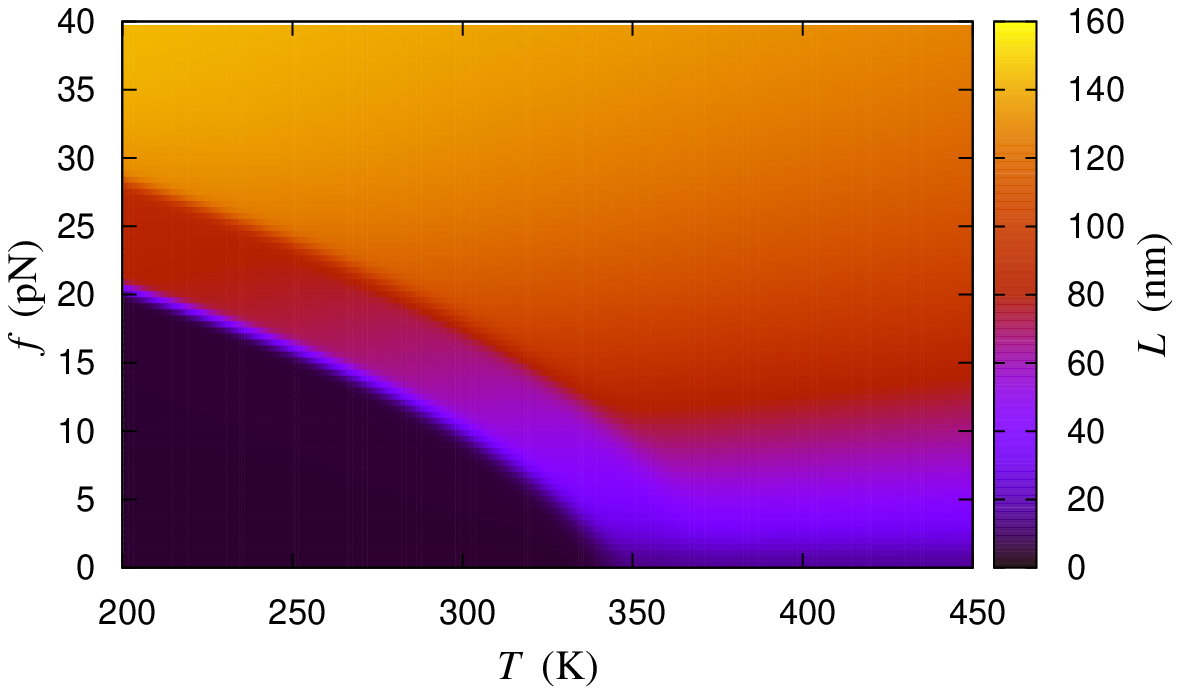}
\caption{(Color online) Force-Temperature phase diagrams of the
  T. thermophila ribozyme. Upper panel: average order parameter $\media
  m$. Lower panel: average end--to--end length $\media
  L$.}\label{phase_grz}
\end{figure}

In fig.~\ref{phase_grz} we plot the $(f,T)$ phase diagram of the 1GRZ
molecule. At variance with the case of the hairpin, this larger
molecule exhibits a wider transition region, reflecting the presence
of intermediate states along the unfolding pathway, as will be
discussed in detail below.

\begin{figure}
\center
\includegraphics[width=8cm]{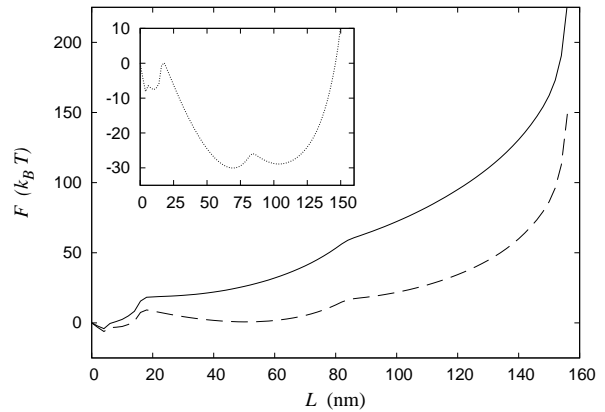}
\caption{Free energy landscape $F_0$ of the 1GRZ molecule as
  a function of the molecular elongation $L$, for $T=300$ K (full
  line). Tilted FEL $F(L,f)$ for two values of the
   force $f=8.82$ pN (dashed line) and $f=17.64$
  pN (dotted line, inset). }
\label{free_grz}
\end{figure}

In fig.~\ref{free_grz}, we plot the unperturbed FEL $F_0$ as a
function of the end--to--end length $L$ at $T=300$ K. In the same
figure, we plot the tilted FEL $F(L,f)$ for two values of the external
force $f=8.82, \, 17.64$ pN. From the phase diagram in
fig.~\ref{phase_grz} one sees that for $f= 17.64$ and $T=300$ K, the
average end--to--end length is half of its maximum value
$L_{\mathrm{max}}\simeq155$ nm. The FEL at $f=17.64$ pN exhibits two
major and one minor energy wells, indicating the coexistence of three
states characterized by three different values of $L$ for this force.

In order to study the mechanical unfolding of the ribozyme, we
consider here the experimental protocol where the external force is
applied by tethering the molecule to a colloidal particle trapped in
an optical trap: $f(t)=k (X(t)-L(t))$, where $L(t)$ is the end-to end
length of the molecule at time $t$, while $k$ and $X(t)=r\cdot t$ are
the stiffness and the center of the trap, respectively.  This
experimental setup corresponds to that used in ref.~\cite{bus2}, which
we will use as a reference to compare our results.

In ref.~\cite{AlbPrl08}, the present model was used to trace the
state of SSEs of a protein. Similarly, one can monitor the unfolding
of a single SSE of 1GRZ by considering a suitable order parameter for
each SSE. Here we choose the fraction of native contacts within an
element.  The order parameter of the SSE Pn will be defined as
$\phi_{\mathrm{Pn}}=\sum'_{ij} \Delta_{ij} \prod_{k=i}^j
m_k/N_{\mathrm{Pn}}$, where the prime means that $i$ and $j$ run over
those bases belonging to Pn, and $N_{\mathrm{Pn}} = \sum'_{ij}
\Delta_{ij}$ is the total number of native contacts in Pn. This
approach allows us to trace the current state of each SSE. The
unfolding time of a given SSE is defined as the time at which the
corresponding order parameter crosses a given threshold $\phi_u=1/3$
for the first time \cite{AlbPrl08}.

\begin{figure}
\center
%\psfrag{ }[ct][ct][1.]{ }
\includegraphics[width=8cm]{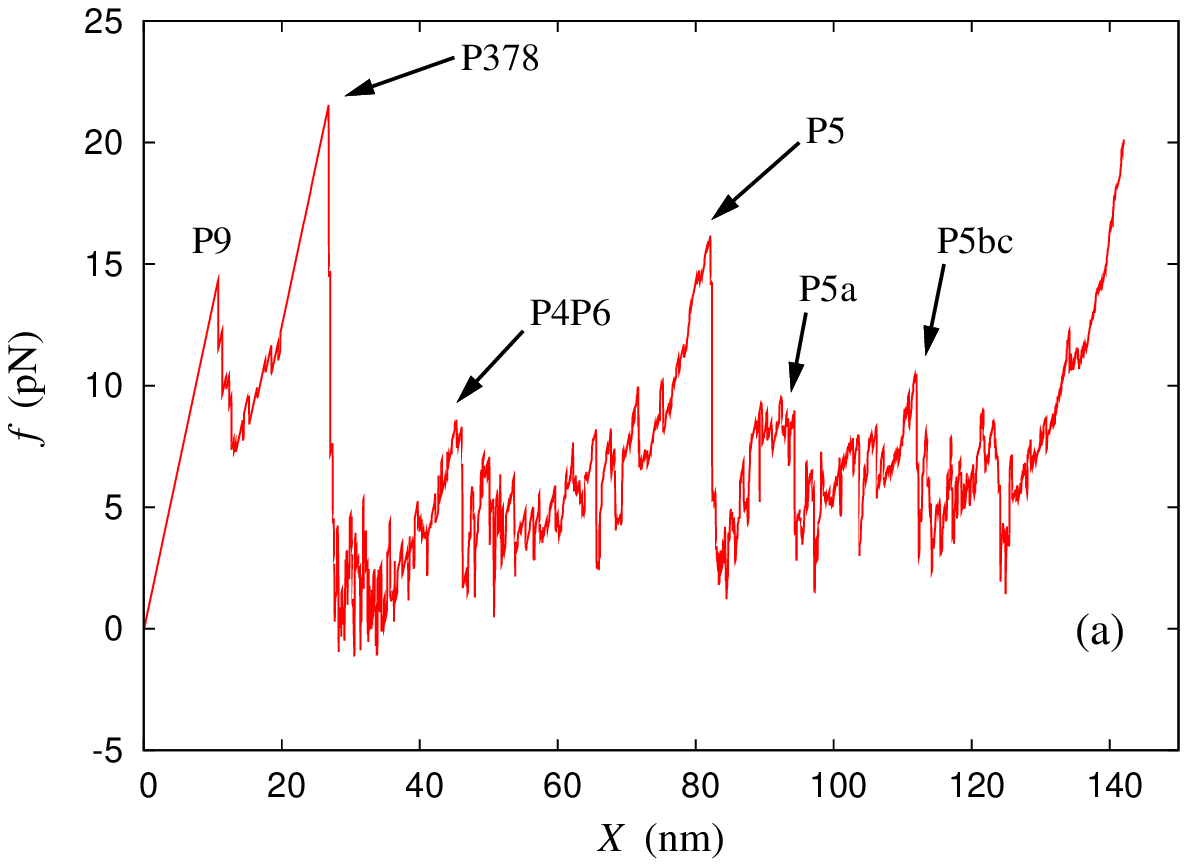}
\includegraphics[width=8cm]{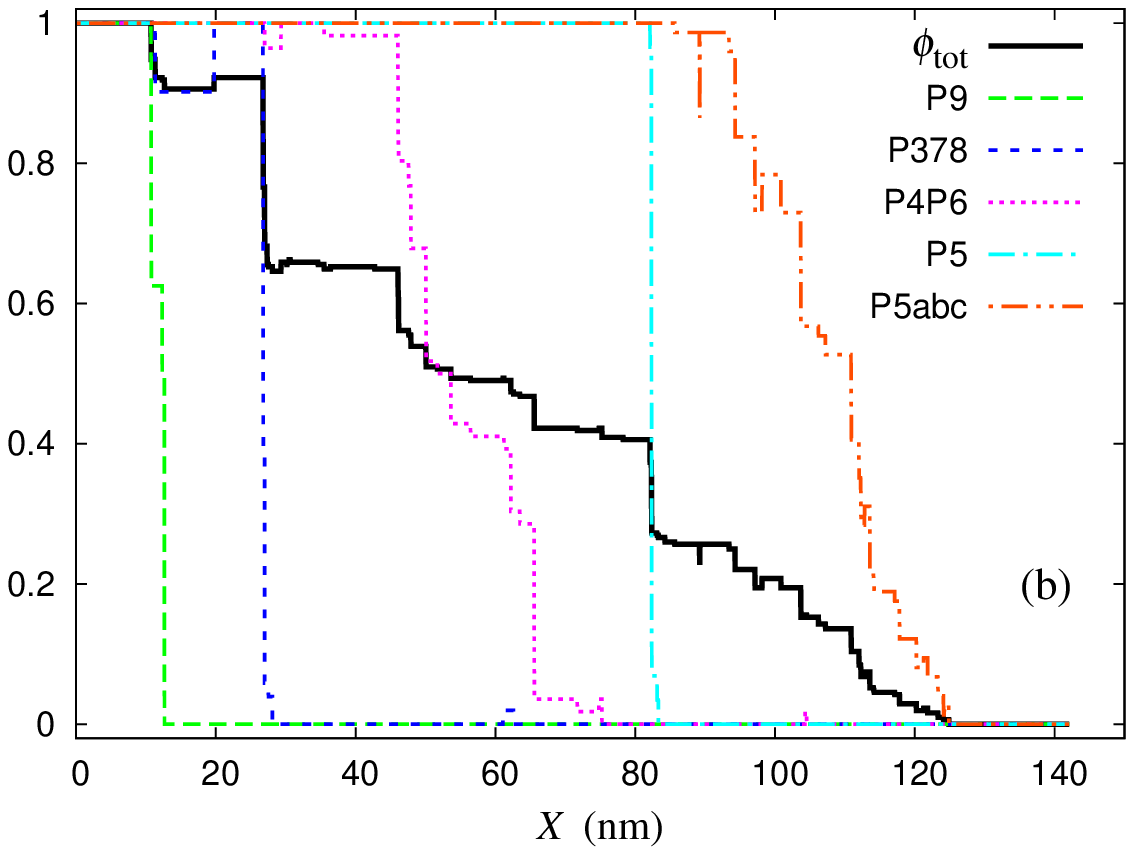}
\caption{(Color online) Typical unfolding trajectory of the 1GRZ molecule. (a) force as a
  function of the position of the ``optical trap'' center $X$. (b)
  order parameter of the whole system and of the single SSEs as
  functions of $X$.}
\label{traj_grz}
\end{figure}

In order to find the typical unfolding pathways, we consider 1000
trajectories, simulated with a standard Monte Carlo algorithm
\cite{AlbPrlJCP07}, where the trap stiffness and the velocity
take the values $k=13$ pN/nm, and $r=0.36$ nm/(MC Step).
Time is a discrete variable counting the number of MC steps.  A
typical trajectory is plotted in fig.~\ref{traj_grz}. In the
force--extension curve unfolding peaks can be observed. By comparing
the position of the peaks with the drops in the SSEs order parameters
in fig.~\ref{traj_grz}(b), we can associate peaks in
fig.~\ref{traj_grz}(a) to the unfolding of SSEs. 
The equilibrium behaviour of these order parameters is reported and discussed in \cite{append}.
The unfolding
pathways corresponding to the 1000 trajectories can be easily clustered
into two big sets. The first set (622 trajectories) corresponds to the
pathway P9 \arr P34678 \arr P5 \arr P5a \arr P5bc. This means that P9
is the first SSE to unfold, followed by the SSEs P3, P4, P6, P7 and P8
with no definite order among them, and so on. This is consistent with
the experimental pathway \cite{bus2} P9 \arr P378 \arr P46 \arr P5
\arr P5abc, except for lumping together P378 and P46 and for splitting
P5abc. This set of trajectories can be analyzed in more detail,
looking for finer subdivisions, and one finds that 429 out of these 622
trajectories correspond to the pathway P9 \arr P78 \arr P3 \arr P46
\arr P5 \arr P5a \arr P5bc. Even at this finer level our results are
consistent with the experimental ones \cite{bus2}, and we can also
predict a definite order of unfolding events within the domains P378
and P5abc. Moreover, we find an alternative, less probable (355
trajectories out of 1000) pathway, P9 \arr P3 \arr P46 \arr P5 \arr
P5a \arr P5bc \arr P78, where P78 are the last domains to unfold.
The order of the unfolding events does not appear to be affected by moderate variations of  the threshold $\delta$ used to obtain the model interaction parameters $\epsilon_{ij}$ \cite{append}.
It is worth to note that such parameters take implicitly into account the
effect of counterions on the stability of the native structure, see same reference.

To summarize, the present model turns out to be able to provide the
equilibrium properties of an RNA hairpin with minimal computational
efforts in comparison with more detailed molecular models.  This
feature allows us to extend our investigation to the equilibrium
properties of a large molecule, namely the Tetrahymena thermophila
ribozyme.  The model is also able to reproduce the experimental
behaviour of the ribozyme under mechanical loading, providing
additional information on the unfolding events at a microscopic level
not accessible to experiments, similarly to what obtained for the
analysis of protein unfolding \cite{AlbPrl08}.  These results clearly
indicate that such a model captures the basic universal features
underlying the mechanical unfolding of biopolymers.

\bibliography{ref_rna}

\begin{thebibliography}{13}
\bibitem[{app()}]{append}
\bibinfo{note}{In this appendix we (1) report the secondary structures of the
  RNA fragments that we have studied, (2) study the equilibrium order
  parameters as functions of the molecular elongation, for comparison with our
  nonequilibrium simulations, (3) report nonequilibrium unfolding simulations
  of molecule 1EOR, and (4) study the effect of the cutoff distance used to
  define contacts on the 1GRZ unfolding nonequilibrium simulations.}
\end{thebibliography}

%\begin{thebibliography}{}

%\bibitem {bus1} J. Liphardt, B. Onoa, S.B. Smith, I. Tinoco, C. Bustamante, Science {\bf 292}, (2001) 733--737.
%\bibitem {bus2} B. Onoa, S. Dumont, J. Liphardt, S.B. Smith, I. Tinoco, C. Bustamante, Science {\bf 299},
%(2003) 1892--1895.
%\bibitem{noi\footnotetext1} A. Imparato, A. Pelizzola, M. Zamparo, {\it Phys. Rev. Lett.}, {\bf 98}: 148102 (2007); A. Imparato, A. Pelizzola, M. Zamparo,  {\it J. Chem. Phys}, {\bf 127}: 145105 (2007).
%\bibitem{noi2}  A. Imparato, A. Pelizzola,  {\it Phys. Rev. Lett.}  {\bf 100}: 158104 (2008).
%\bibitem{thiru1} C. Hyeon, and D. Thirumalai PNAS {\bf 102}, 6789-6794 (2005).
%\bibitem{thiru2} C. Hyeon, and D. Thirumalai Biophys. J. {\bf 92}, 731-743 (2007).
%\bibitem{Felix05}D. Collin {\it et al.}, Nature, 437 (2005) 231-234.
%\bibitem{crooks} G. E. Crooks, Phys. Rev. E {\bf 60}, 2721--2726 (1999).
%\bibitem{nota} The force--ramp is one of the possible experimental setup for the unfolding of biomolecules.
%Different setups are obtained by pulling the molecule with the probe  of an AFM, or by  thethering it to a colloidal particle trapped in a laser trap. In these cases, the free energy of the model molecule considered here can still be calculated exactly\footnotetext by summing a partition function similar to that appearing in eq.~(\ref{def_g}), where the term $-f L$ has to be replaced by the potential associated to the external pulling device, which is usually quadratic.
%\end{thebibliography}

\newpage
\clearpage

\newcommand{\eij}{\epsilon_{ij}}
\newcommand{\dij}{\Delta_{ij}}
\renewcommand{\thefigure}{A.\arabic{figure}}

\begin{widetext}

{\Large \bf
Appendix to ``Equilibrium properties and force-driven unfolding pathways of RNA molecules''}

\section{Molecular structures}
In this section we report the secondary structures of the RNA
fragments we have studied with our model. 

In fig.\ \ref{fig1} we report
the secondary structure of the simple hairpin P5GA (pdb code 1EOR).

In addition, in fig.~\ref{fig2} we report the secondary structure of
the main structured portion of the Tetrahymena thermophila ribozyme
(pdb code 1GRZ). This is the portion we have considered in our work
and goes from base 96 to 331, according to the pdb numbering. In the
figure, the numbers which are adjacent to bases correspond to the pdb
numbering, and the labeling of hairpins follows \cite{golden98}, where
the secondary structure of the whole molecule can be found. 
It is worth to note, that in that work the authors obtained crystals of the ribozyme by freezing 
solutions containing 50 mM of Mg$\mathrm{Cl}_2$. This concentration is compatible with the experimental
conditions of, e.g., \cite{bus1,bus2}.
Thus, the native structure, as given in the protein data base PDB,  is affected by the presence of ions $\mathrm{Mg}^{2+}$, which stabilize the long range, tertiary contacts.
As discussed in the text, in generating our model interaction energies $\eij$, we take into account those atom pairs whose distance is smaller than a threshold distance.
Thus the effect of the Mg+ ions is implicitly taken into account in our model.
It is commonly believed [2,3] that tertiary contacts are bottlenecks for the 
molecular unfolding, which manifest as  peaks in the force-extension curve,
like those in fig.5a of the main text.

\begin{figure}[h]
\center
\includegraphics{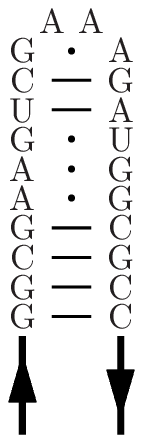}
\vspace{1cm}
\caption{Secondary structure of 1EOR}
\label{fig1}
\end{figure}

\begin{figure}[h]
\center
\includegraphics{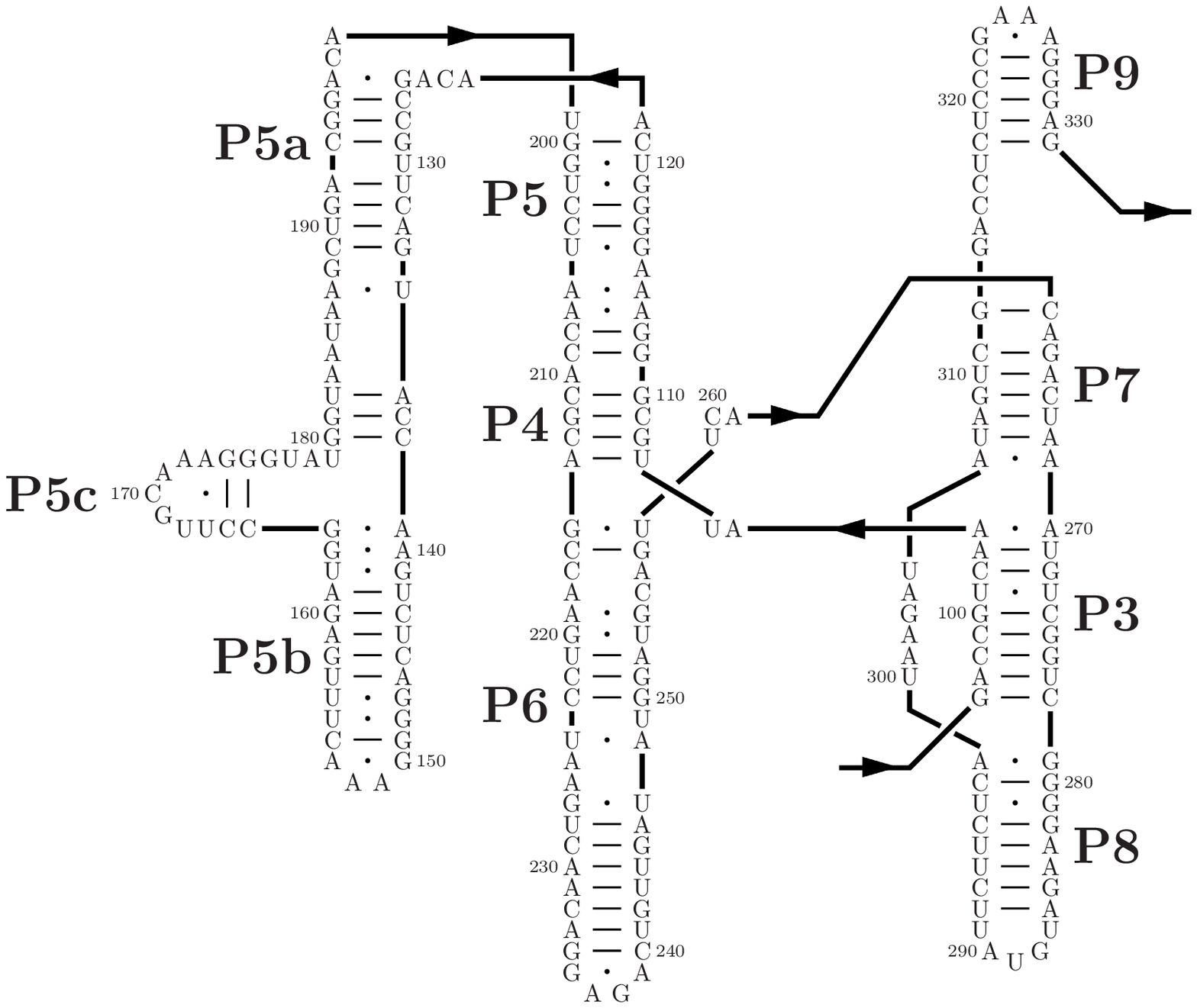}
\caption{Secondary structure of 1GRZ}
\label{fig2}
\end{figure}

\section{Comparison of the out--of--equilibrium unfolding pathways with the equilibrium configurations}
\begin{figure}[h]
\center
\includegraphics[width=8cm]{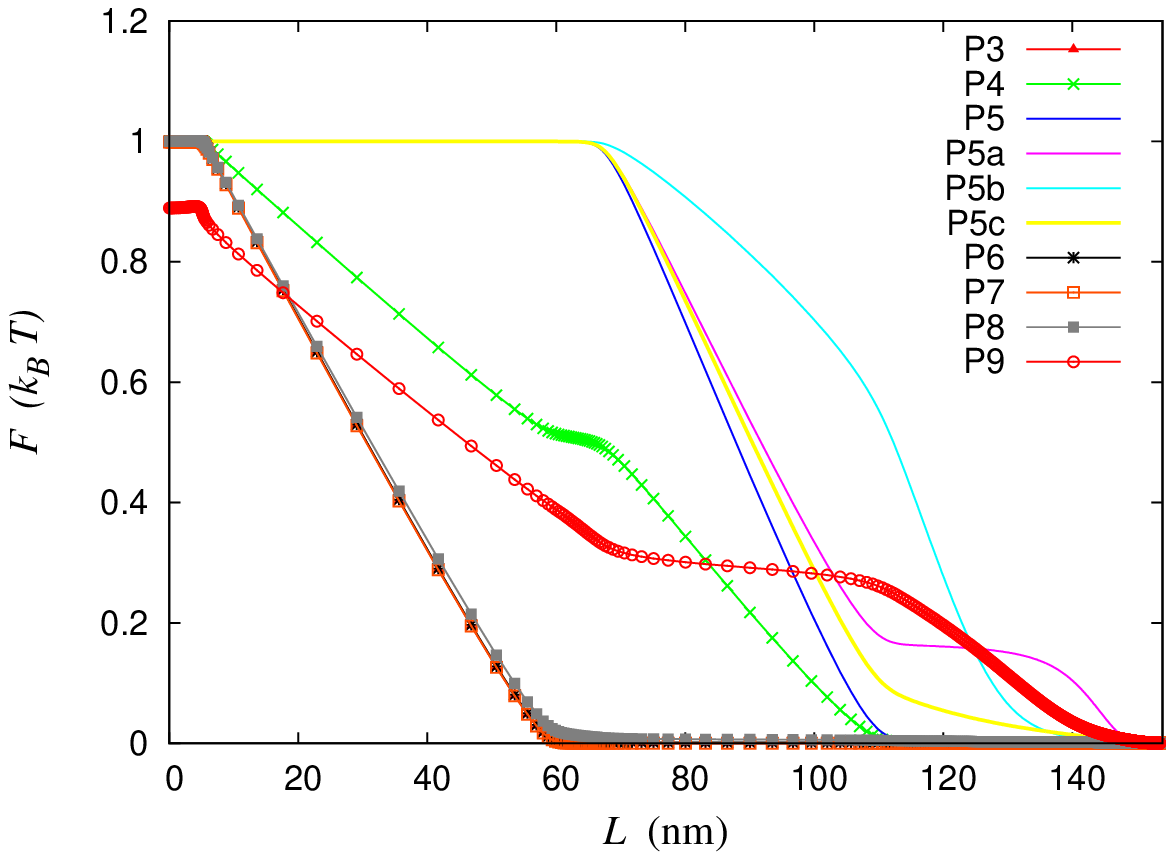}
\caption{Equilibrium order parameter $\phi$ as a function of the equilibrium molecular elongation $L$, at $T=300$ K.}
\label{fig3}
\end{figure}
In fig.~\ref{fig3} the order parameter $\phi$ as a function of the molecular elongation $L$, at $T=300$ K.  Inspection of this curve suggests that at such a temperature, P9 is partially unfolded, and as the molecular elongation increases, the unfolding   of P3, P4,P6, P7, and P8 start at $L=6$ nm, the unfolding of P4 and P9 being slower than the others.
At $L=65$ nm, the P3, P6, P7 and P8 are completely unfolded, while the curves for P4 and P9 exhibits a shoulder. At the same value of the elongation, the SSEs P5, P5a, P5b, and P5c start to unfold too, and for $L\simeq$ 115 nm only a small fraction of P9 and of P5abc is not yet unfolded.
Such a figure has to be compared with fig. 4 and fig. 5 in the main text. The picture emerging is consistent with the typical unfolding pathways
found during the simulations: P9 \arr P34678 \arr P5 \arr P5a \arr P5bc. It is worth to note, however, that in  out-of-equilibrium pulling
manipulations  the unfolding is a stochastic process, and thus the  sequence of the unfolding events may vary from one realization of the process to another. For example in fig 5.b in the main text one clearly distinguish the unfolding events of P378 and P4P6, while in the equilibrium
picture,  fig.~\ref{fig3}, the unfolding of those elements occurs at almost the same length.
Similarly, the characteristic lengths emerging from analysis of fig.~\ref{fig3}, i.e. $L=6,\, 65$ nm correspond to  
minima in the energy landscape plotted in the inset of fig. 4 in the main text, suggesting that those minima correspond to the typical
configurations  of the molecule immediately before the unfolding of a specific group of SSEs.

\section{Supplementary simulations}

{\it 1EOR}

Here, we describe  unfolding simulations of the 1EOR molecule.
We consider first the constant force setup (force clamp): a constant force
$f=15.4$ pN is applied to the molecule, and its length is traced as a function
of time. A typical trajectory is plotted in fig.~\ref{fc_eor}. Inspection of this figure
clearly indicates that the molecule  hops back
and forth between two states characterized by $L\simeq 4$ nm and $L\simeq10$ nm, which correspond to the two minima in the energy landscape of the molecule,
see inset of fig. 2 in the main text.

\begin{figure}[h]
\center
\includegraphics[width=8cm]{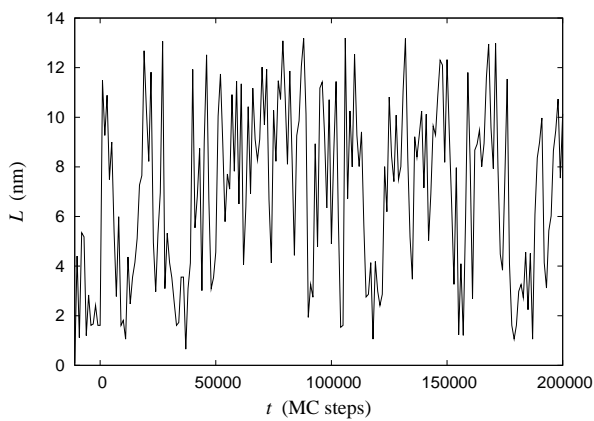}
\caption{Typical unfolding trajectory of the 1EOR molecule at $T=300$ K, under constant force.}
\label{fc_eor}
\end{figure}

We now consider the dynamic-loading set up.
As in the main text, the force applied changes as a function of time as $f(t)=k (X(t)-L(t))$, where $L(t)$ is the end-to end
length of the molecule at time $t$, while $k$ and $X(t)=r\cdot t$ are
the stiffness and the center of the external potential, respectively.
We take the values $k=13$ pN/nm, and $r=1$ nm/(MC step). 
We find that the molecule unfolds exhibiting a peak in the force-elongation curve, at $f\simeq 12$ pN, in very good agreement with ref.~\cite{bus1}.

\begin{figure}[h]
\center
\includegraphics[width=8cm]{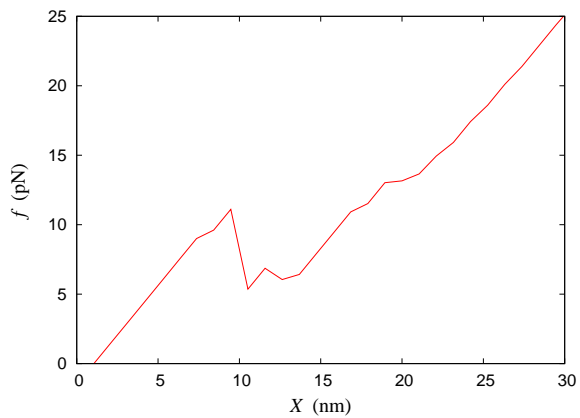}
\caption{Typical unfolding trajectory of the 1EOR molecule at $T=300$ K, under time-varying force. $X$ is the position of the center of the external quadratic potential.}
\label{fig_eor_a}
\end{figure}

{\it 1GRZ}

As stated in the main text, in the present RNA model, bases $i$ and $j+1$ are
considered to be in contact if at least two atoms, one from each
base, are closer than $\delta=4$ \AA. We now study the possible effects  of  the threshold $\delta$ on the unfolding pathway, and in particular on 
the force-elongation curve.

We consider the values $\delta=3.5,\, 4.5,\, 8$ \AA, and run unfolding simulations as those described in the main text. 
Typical trajectories are plotted in fig.~\ref{fig4}: inspection of these figures suggests that increasing $\delta$ has no net effect
on the unfolding pathway, the typical unfolding sequence being that of fig. 5 in the main text.
For $\delta=3.5$ however, the peaks in the force-extension curve appear to be flattened out, as one would expect, because of the reduced number and intensity of contact interactions. 

\begin{figure}[h]
\center
\includegraphics[width=8cm]{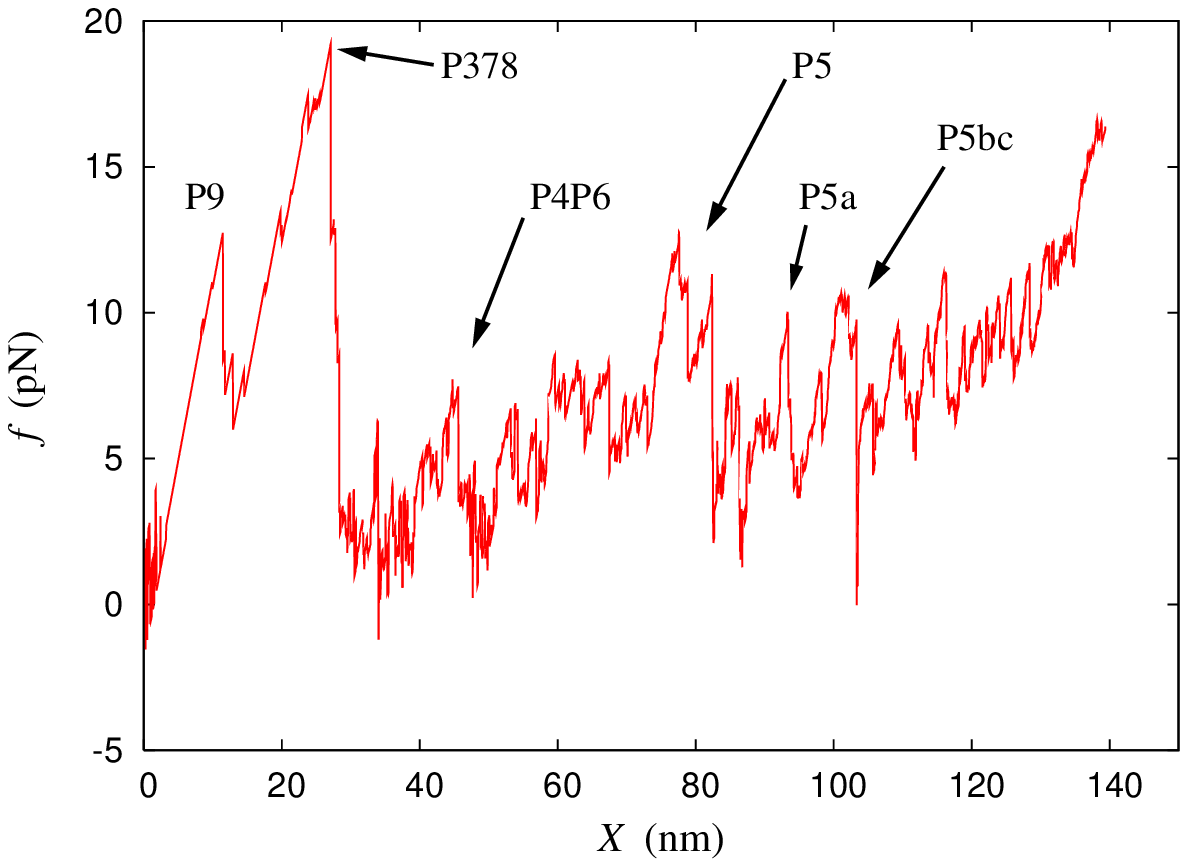}
\includegraphics[width=8cm]{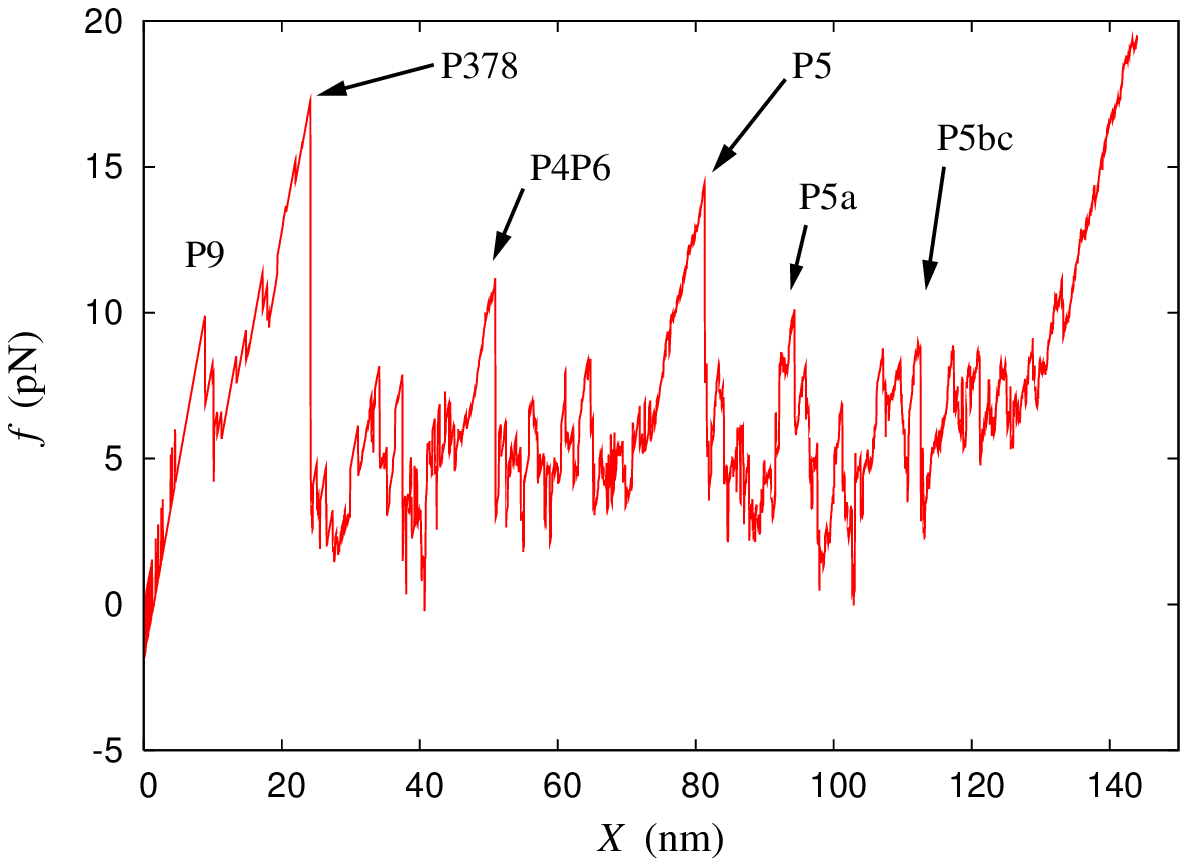}
\includegraphics[width=8cm]{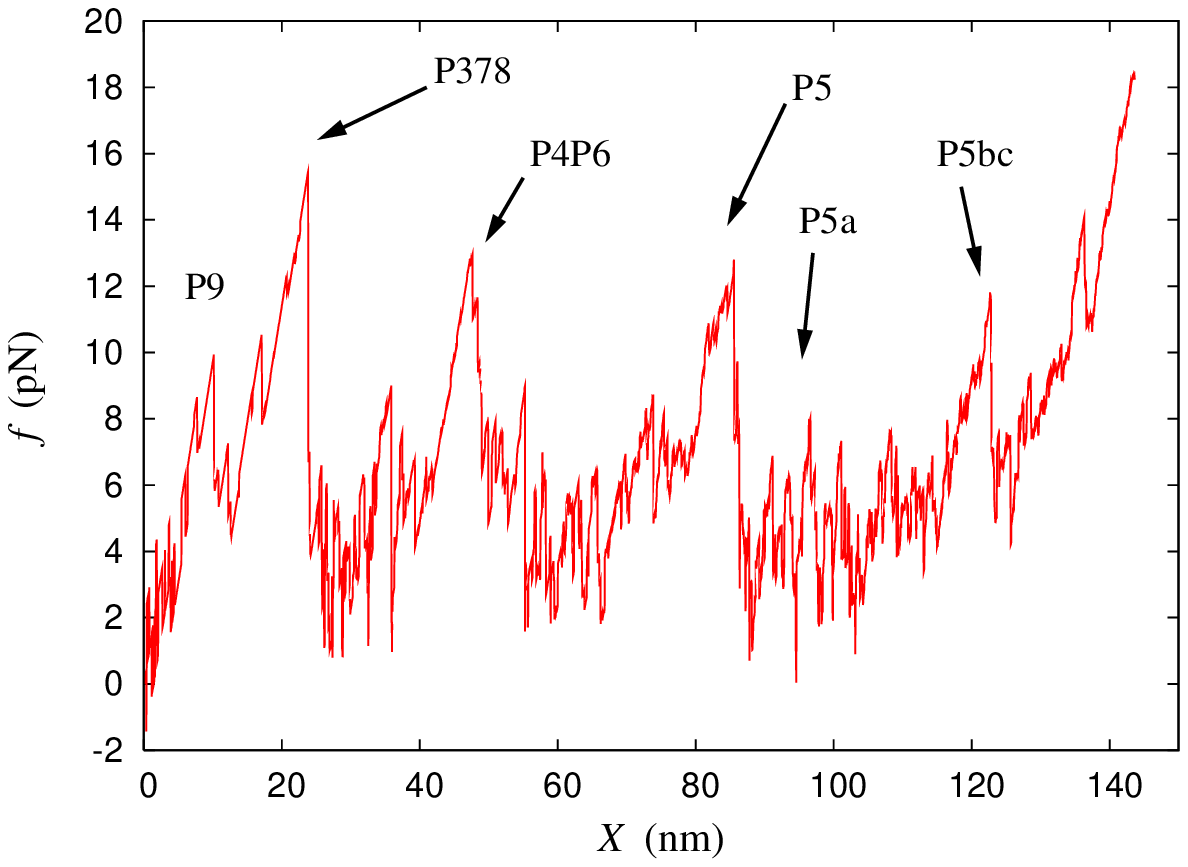}
\caption{Typical unfolding trajectory of the 1GRZ molecule with modified interaction parameters $\eij$. The parameters have been obtained with $\delta=3.5$ (a) , 4.5 (b) and  8 (c) \AA .}
\label{fig4}
\end{figure}

\end{widetext}

\end{document}